\documentclass{PoS}

\let\Otemize =\itemize
\let\Onumerate =\enumerate
\let\Oescription =\description
\def\Nospacing{\itemsep=0pt\topsep=0pt\partopsep=0pt\parskip=0pt\parsep=0pt}
\def\Topspac{\vspace{-0.5\baselineskip}}
\def\Botspac{\vspace{-0.5\baselineskip}}
\newenvironment{Itemize}{\Topspac\Otemize\Nospacing}{\endlist\Botspac}

\newcommand{\sqrtsNN}{\sqrt{s_{\scriptscriptstyle \rm NN}}}
\newcommand{\av}[1]{\left\langle #1 \right\rangle}

\newcommand{\gev}{\mathrm{GeV}}
\newcommand{\tev}{\mathrm{TeV}}

\newcommand{\mum}{\mathrm{\mu m}}

\renewcommand{\d}{\mathrm{d}}

\newcommand{\pt}{p_{\rm T}}

\newcommand{\dEdx}{{\rm d}E/{\rm d}x}
\newcommand{\RAA}{R_{\rm AA}}

\title{Suppression of high-$\pt$ heavy-flavour particles in Pb--Pb collisions at the LHC, measured with the ALICE detector}

\ShortTitle{Suppression of high-$\pt$ heavy-flavour particles in Pb--Pb collisions at the LHC, measured at ALICE}

\author{\speaker{Andrea Dainese}, for the ALICE Collaboration\\
       INFN - Sezione di Padova, Padova, Italy\\
       E-mail: \email{andrea.dainese@pd.infn.it}}

\abstract{
The ALICE experiment studies nucleus--nucleus collisions at the LHC in order to 
investigate the properties of QCD matter at extreme energy densities. 
The measurement of open charm and open beauty production allows one to 
probe the mechanisms of heavy-quark propagation, energy loss and
hadronization in the hot and dense medium formed in high-energy 
nucleus-nucleus collisions. In particular, in-medium energy loss is 
predicted to be different for massless partons (light quarks and gluons) 
and heavy quarks at moderate momentum. 
The ALICE apparatus allows us to measure open heavy-flavour particles
in several decay channels and with a wide phase-space coverage.

We present the results on the nuclear modification factors for heavy-flavour
particle production in Pb--Pb collisions at $\sqrtsNN=2.76$~TeV.
Using proton--proton and lead--lead collision samples at $\sqrt{s}=2.76$ and 7~TeV and at 
$\sqrtsNN=2.76$~TeV, respectively,  
nuclear modification factors $R_{\rm AA}(p_{\rm T})$ were measured for D mesons
at central rapidity (via displaced decay vertex reconstruction), and for electrons and muons from heavy-flavour decays, 
at central and forward rapidity, respectively.
A large suppression is observed, by a factor 2.5--4 in central Pb--Pb collisions with respect to 
the pp reference, in the high $p_{\rm T}$ region, indicating a strong in-medium energy loss of heavy quarks.
}

\FullConference{36th International Conference on High Energy Physics,\\
		July 4-11, 2012\\
		Melbourne, Australia}

\begin{document}

\section{Introduction}

The ALICE experiment~\cite{aliceJINST} studies proton--proton, proton--nucleus and nucleus--nucleus
collisions at the Large Hadron Collider (LHC), with the
main goal of investigating the properties of the high-density, colour-deconfined,
state of strongly-interacting matter that is expected to be formed
in Pb--Pb collisions. 
Pb--Pb data were collected in November 2010 and in November 2011 at 
a centre-of-mass energy $\sqrtsNN=2.76~\tev$ per nucleon--nucleon collision.

Heavy-flavour particles, abundantly produced at LHC energies,
are regarded as effective probes of the conditions of the system (medium) formed 
in nucleus--nucleus collisions. In particular, 
they should be sensitive to the energy density,
through the mechanism of in-medium energy loss of heavy quarks.
The nuclear modification factor $R_{\rm AA}$ of particle $\pt$ distributions
is well-established as a sensitive observable 
for the study of the interaction of hard partons 
with the medium. This factor is defined as 
the ratio of the $\pt$ spectrum measured in nucleus--nucleus (AA)
to that expected on the basis of the proton--proton spectrum scaled 
by the number $N_{\rm coll}$ of binary 
nucleon--nucleon collisions in the nucleus--nucleus collision: 
$R_{\rm AA}(\pt)=
(1/\av{T_{\rm AA}}) \cdot 
(\d N_{\rm AA}/\d\pt)/ (\d\sigma_{\rm pp}/\d\pt)$,
where the AA spectrum corresponds to a given collision-centrality class and 
$\av{T_{\rm AA}}$, proportional to $\av{N_{\rm coll}}$, is the average nuclear overlap function resulting for that 
centrality class from the 
Glauber model of the collisions geometry~\cite{glauber}.
Due to the 
QCD nature of parton energy loss, quarks are predicted to lose less
energy than gluons (that have a higher colour charge) and, in addition, 
the `dead-cone effect' and other mechanisms are expected to introduce a mass-dependence 
in the coupling of hard partons with the medium constituents~\cite{dk,asw,whdg,rapp}. 
Therefore, one should observe a pattern
of gradually decreasing $\RAA$ suppression when going from the mostly 
gluon-originated
light-flavour hadrons (e.g. pions) 
to D and to B mesons~\cite{whdg,adsw}: 
$\RAA^\pi<\RAA^{\rm D}<\RAA^{\rm B}$. 
The measurement and comparison of these different medium probes 
provide a unique test of the 
colour-charge and mass dependence of parton energy loss.

The ALICE experiment was designed to detect heavy-flavour hadrons
over a wide phase-space coverage and in various decay channels, as we will 
discuss in section~\ref{sec:hf}.
The production of heavy-flavour probes was measured in proton--proton collisions at 
$\sqrt{s}=2.76$ and $7~\tev$ and compared to perturbative QCD (pQCD) predictions. These results,
summarized in section~\ref{sec:pp}, provide a well-defined, calibrated, reference for 
$\RAA$ measurements. The Pb--Pb analysis results and 
the nuclear modification factors of D mesons, and of 
electrons and muons from heavy-flavour decays are presented in 
section~\ref{sec:pbpb}, along with a comparison with other measurements and with model calculations. 

\section{Heavy-flavour production measurements in ALICE}
\label{sec:hf}

The heavy-flavour detection capability of the ALICE detector,
described in detail in~\cite{aliceJINST}, is mainly provided by:
\begin{Itemize}
\item Tracking system; the silicon Inner Tracking System (ITS) and 
the Time Projection Chamber (TPC),
embedded in a magnetic field of $0.5$~T, provide track reconstruction in 
the pseudo-rapidity range $|\eta|<0.9$ 
with a momentum resolution better than
4\% for $\pt<20~\gev/c$ 
and a transverse impact parameter\footnote{The transverse impact parameter,
$d_0$, is defined as the distance of closest approach of the track to the 
interaction vertex, in the plane transverse to the beam direction.} 
resolution better than 
$65~\mum$ for $\pt>1~\gev/c$ with a high-$\pt$ value of $20~\mum$~\cite{aliceDRAA}.
\item Particle identification system; charged hadrons are separated via 
their specific energy deposit $\dEdx$ in the TPC and via time-of-flight measurement in the 
Time Of Flight (TOF) detector; 
electrons are identified at low $\pt$ ($<6~\gev/c$) via TPC $\dEdx$ and 
time-of-flight, and at intermediate and high $\pt$ ($>2~\gev/c$) in the dedicated Transition Radiation Detector (TRD) and
in the Electromagnetic Calorimeter (EMCal); 
muons are identified in the muon 
spectrometer covering the pseudo-rapidity range $-4<\eta<-2.5$. 
\end{Itemize}
The main detection modes are listed below.
\begin{Itemize}
\item Charm hadronic decays in $|y|<0.5$ using displaced vertex
identification: 
$\rm D^0 \to K^-\pi^+$,  
$\rm D^+ \to K^-\pi^+\pi^+$, $\rm D^{*+}\to D^0\pi^+$, and $\rm D_s^+ \to K^-K^+\pi^+$. 
\item Leptons from heavy-flavour decays: 
inclusive ${\rm D/B\to e}+X$ in $|y_{\rm e}|<0.8$ and ${\rm D/B\to\mu}+X$ in $2.5<y_{\mu}<4$; 
${\rm B\to e}+X$ using displaced electron identification and ${\rm B\to J/\psi\,(\to e^+e^-)}+X$ (only in pp for the moment).
\end{Itemize} 

The detection strategy for D mesons at central rapidity
is based, for both pp and Pb--Pb collisions,  
on the selection of displaced-vertex topologies, i.e. separation of tracks
from the secondary vertex from those from the primary vertex, 
large decay length (normalized to its estimated uncertainty),
and good alignment between the reconstructed D meson momentum 
and flight-line. 
The identification of the charged kaon in the TPC and TOF detectors helps to further reduce the
background at low $\pt$. 
An invariant-mass analysis is then used to extract the raw signal 
yield, to be then corrected for detector acceptance and 
for PID, selection and reconstruction efficiency, evaluated from a detailed detector simulation. 
The contamination of D mesons from B meson
decays is estimated to be of about 15\%, using the beauty production cross
section predicted by the FONLL (fixed-order next-to-leading log) 
calculation~\cite{fonllpriv} and the detector simulation, 
and it is subtracted from the measured raw $\pt$ spectrum, before applying the efficiency 
corrections.
The systematic uncertainty related to the FONLL theoretical uncertainty is correlated in the 
pp and Pb--Pb spectra, thus it partly cancels in the $\RAA$ ratio. An additional source of 
systematic uncertainty arises from the unknown nuclear modification of
B meson production. To estimate this uncertainty, the prompt D meson $\RAA$
was evaluated for the range of hypotheses $0.3<\RAA^{\rm B}/\RAA^{\rm D}<3$~\cite{aliceDRAA}.

At central rapidity, heavy-flavour production is measured also using semi-electronic decays.
The basis of this measurement is a robust electron identification. 
This uses the signals of four different detectors: TPC, TOF, EMCal and TRD (only in pp collisions for the moment).
The residual pion contamination in the electron sample is measured with fits to the TPC $\dEdx$ and EMCal $E/p$ distributions
in momentum slices, and subtracted.
The $\pt$-differential cross section of electrons from 
charm and beauty particle decays is obtained by subtracting from the 
efficiency-corrected inclusive spectrum the components of background electrons. 
The main components are 
electrons from Dalitz decays of light-flavour hadrons (mainly
    $\pi^0$ and $\eta$, in addition to $\rho$, $\omega$, and $\phi$), 
   $\gamma$ conversions in the beam pipe and innermost pixel layer, and electrons from J/$\psi$ decays.
The latter is estimated with simulations using the measured J/$\psi$ production in pp collisions and 
its nuclear modification factor in Pb--Pb collisions.
The first two components are estimated using two different methods: for the pp analysis, a simulation ``cocktail'' with 
measured hadron cross sections as input~\cite{hfepp7}; for the Pb--Pb analysis, the direct reconstruction of the 
two-track vertices of internal and in-material photon conversions.

Heavy-flavour production at forward rapidity is measured using the single-muon $\pt$ 
distribution.
The extraction of the heavy-flavour contribution from the single muon spectra requires the subtraction of three main sources of background: muons from the decay-in-flight of light hadrons (decay muons); muons from the decay of hadrons produced in the interaction with the front absorber (secondary muons); hadrons that punch through the front absorber.
The last contribution, about 20\% for $\pt>2~\gev/c$, can be efficiently rejected by requiring the matching of the reconstructed tracks with the tracks in the trigger system. Due to the lower mass of the parent particles, the
light-hadron decay muons have a softer transverse momentum than the heavy-flavour muons, 
and dominate the low-$\pt$ region. In the pp analysis this background is subtracted using
simulations, as detailed in~\cite{hfmpp7}. In the Pb--Pb analysis, instead, it is extrapolated from the 
$\pi$ and K spectra measured at central rapidity ~\cite{hfmraa}. 

The results that we present are obtained from data recorded 
with minimum-bias, single muon and EMCal trigger selections during the Pb--Pb runs of 
2010 (D mesons~\cite{aliceDRAA} and muons~\cite{hfmraa}) 
and 2011 (electron preliminary data). 
Pb--Pb collision-centrality classes
are defined in terms of percentiles of the distribution of 
the sum of amplitudes in the VZERO scintillator detectors, located in the forward and backward pseudo-rapidity regions
of the experiment.

\section{Reference measurements in pp collisions at $\sqrt{s}=2.76$ and $7~\rm TeV$}
\label{sec:pp}

The $\rm D^0$, $\rm D^+$, and $\rm D^{*+}$ $\pt$-differential production 
cross sections in $|y|<0.5$ were measured in pp collisions at $\sqrt{s}=2.76$~\cite{Dpp2.76} and $7~\rm TeV$~\cite{Dpp7},
using data samples with integrated luminosities of 1.1 and 5.0~nb$^{-1}$ and covering the $\pt$ intervals 1--24 and 1--12~$\gev/c$, respectively.

The heavy-flavour decay electron cross section in $|y|<0.5$ was measured in pp collisions at $\sqrt{s}=7~\tev$~\cite{hfepp7} 
in the range 0.5--8~$\gev/c$ with an integrated luminosity of 2.6~nb$^{-1}$.
At forward rapidity $2.5<y<4$, the heavy-flavour decay muon cross section was measured in pp collisions at $\sqrt{s}=2.76$~\cite{hfmraa} and $7~\rm TeV$~\cite{hfmpp7},
using data samples with integrated luminosities of 19.0 and 16.5~nb$^{-1}$ and covering the $\pt$ intervals 2--10 and 2--12~$\gev/c$, respectively.

All these measurements are described within uncertainties by theoretical predictions based on pQCD calculations. For the electron and muon spectra, the 
FONLL calculation indicates that beauty decays are the dominant contribution for $\pt$ larger than about 5--$7~\gev/c$.

The reference pp cross sections used for the determination of the nuclear 
modification factors $\RAA$ in Pb--Pb collisions at $\sqrtsNN=2.76~\tev$ were defined as follows:
\begin{Itemize}
\item for D mesons, the cross section measured at $\sqrt{s}=7~\tev$, scaled to $2.76~\tev$;
\item for heavy-flavour decay electrons, the cross section measured at $\sqrt{s}=7~\tev$, scaled to $2.76~\tev$ ($\pt<8~\gev/c$), and the FONLL cross section with its theoretical uncertainty ($\pt>8~\gev/c$);
\item for heavy-flavour decay muons, the cross section measured at $\sqrt{s}=2.76~\tev$ (which has the same precision as at 7~TeV).
\end{Itemize}

The scaling factors for D mesons and electrons
were defined as the ratios of the cross sections from 
the FONLL pQCD calculation at $2.76$ and $7~\tev$~\cite{scaling}.
The scaled D meson cross sections were found to be consistent with those measured, 
though with limited precision of 20--25\%   
in pp collisions at $\sqrt{s}=2.76~\tev$~\cite{Dpp2.76}.

\section{Heavy-flavour nuclear modification factors in Pb--Pb at $\sqrt{s_{\scriptscriptstyle\rm NN}}=2.76~\rm TeV$}
\label{sec:pbpb}

\begin{figure}[!t]
  \begin{center}
\includegraphics[width=0.48\textwidth]{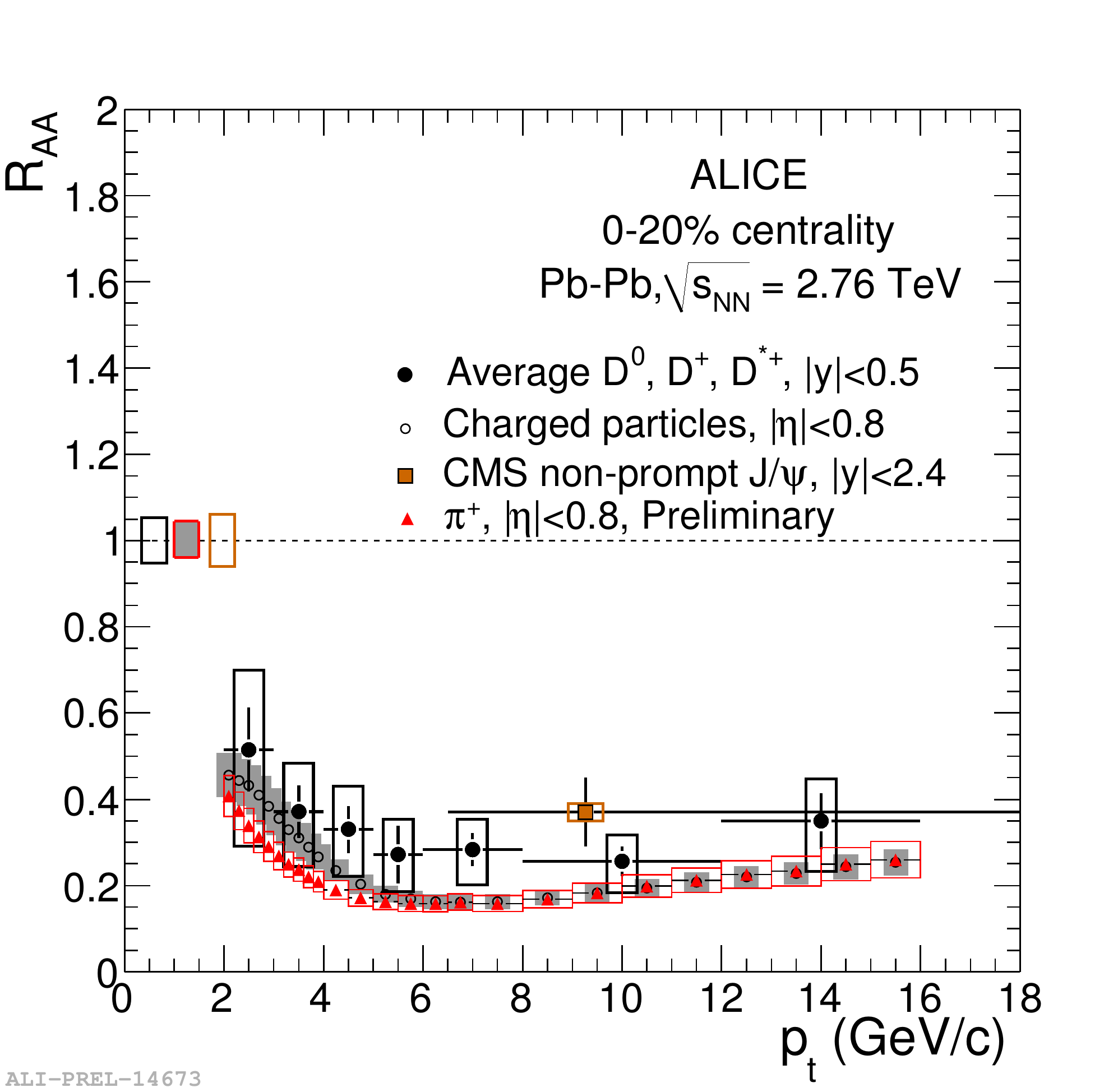}
  \includegraphics[width=0.51\textwidth]{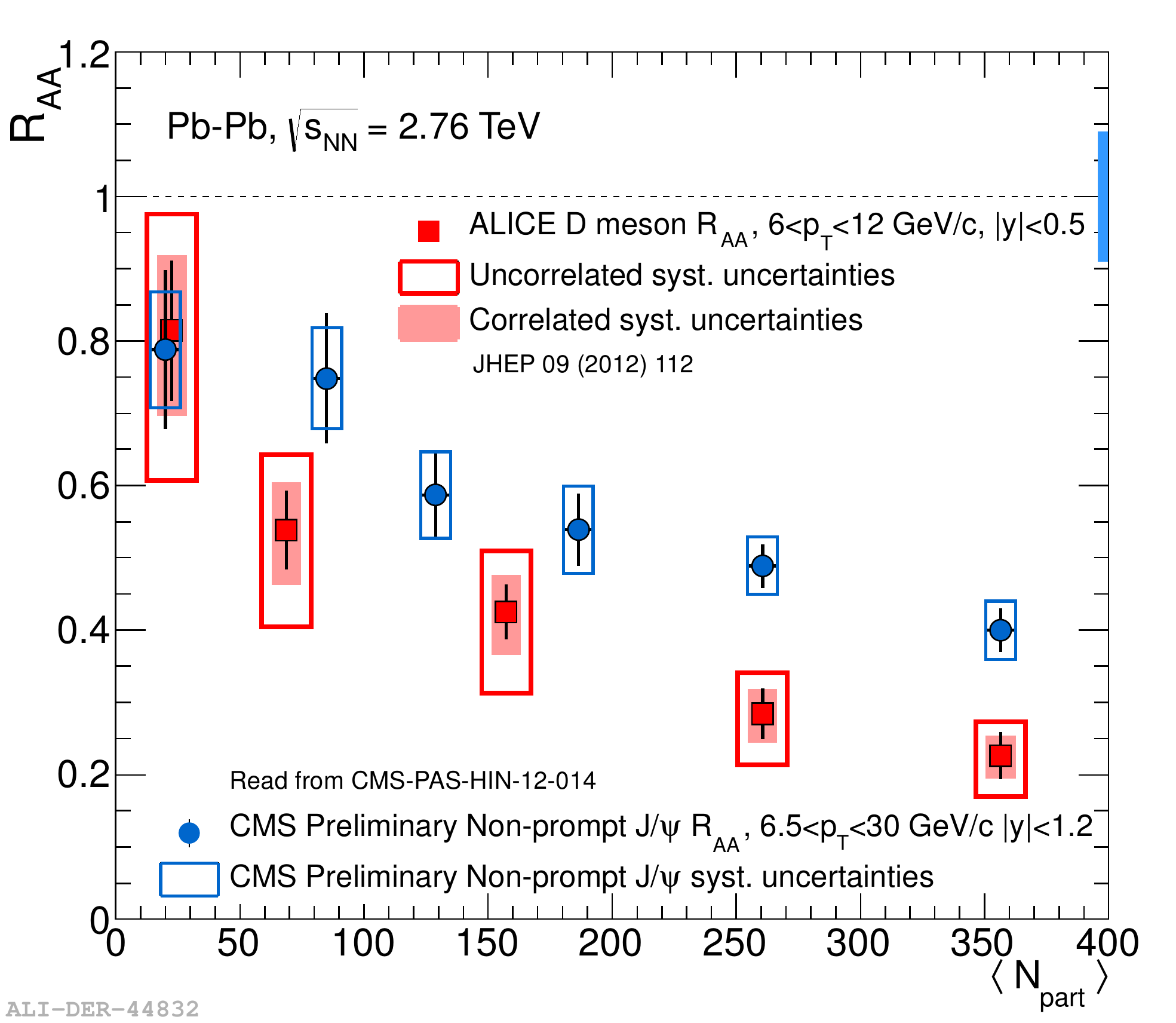}
  \caption{Left: Average $\RAA$ of D mesons in the 0--20\% centrality class~\cite{aliceDRAA}, compared to  
  that of charged particles~\cite{RAAcharged}, $\pi^+$, and non-prompt $\rm J/\psi$
  from B decays~\cite{CMSquarkonia} in the same centrality class. 
  Right: Average $\RAA$ of D mesons as a function of centrality~\cite{aliceDRAA} (average number of nucleons participating in the collision, from the Glauber model~\cite{glauber}, in each centrality class), compared to that of non-prompt $\rm J/\psi$
  from B decays~\cite{CMSpashin12014}. 
  }
\label{fig:DRAA}
\end{center}
\end{figure}

The nuclear modification factor of prompt D mesons
(average of $\rm D^0$, $\rm D^+$ and $\rm D^{*+}$)
in Pb--Pb collisions is shown in figure~\ref{fig:DRAA}: as a function of $\pt$ in central collisions (left) and 
as a function of centrality for $6<\pt<12~\gev/c$ (right)~\cite{aliceDRAA}. 
A strong suppression 
is observed in central collisions, reaching a factor about 4 for $\pt>5~\gev/c$. 
$\RAA$ is compatible with 
that of charged pions and charged particles~\cite{RAAcharged}, although there seems to be a tendency for
$\RAA^{\rm D}>\RAA^\pi$ at low $\pt$. 
In the figure, the nuclear modification of non-prompt J/$\psi$ from B decays, measured by the CMS Collaboration~\cite{CMSquarkonia,CMSpashin12014},
is also shown. The comparison with D mesons indicates a different suppression for charm and beauty hadrons in central collisions, 
consistent with the expectation $\RAA^{\rm D}<\RAA^{\rm B}$. However, the different $\pt$ and rapidity ranges still prevent from a clear conclusion
on this point.

\begin{figure}[!t]
  \begin{center}
\includegraphics[width=0.5\textwidth]{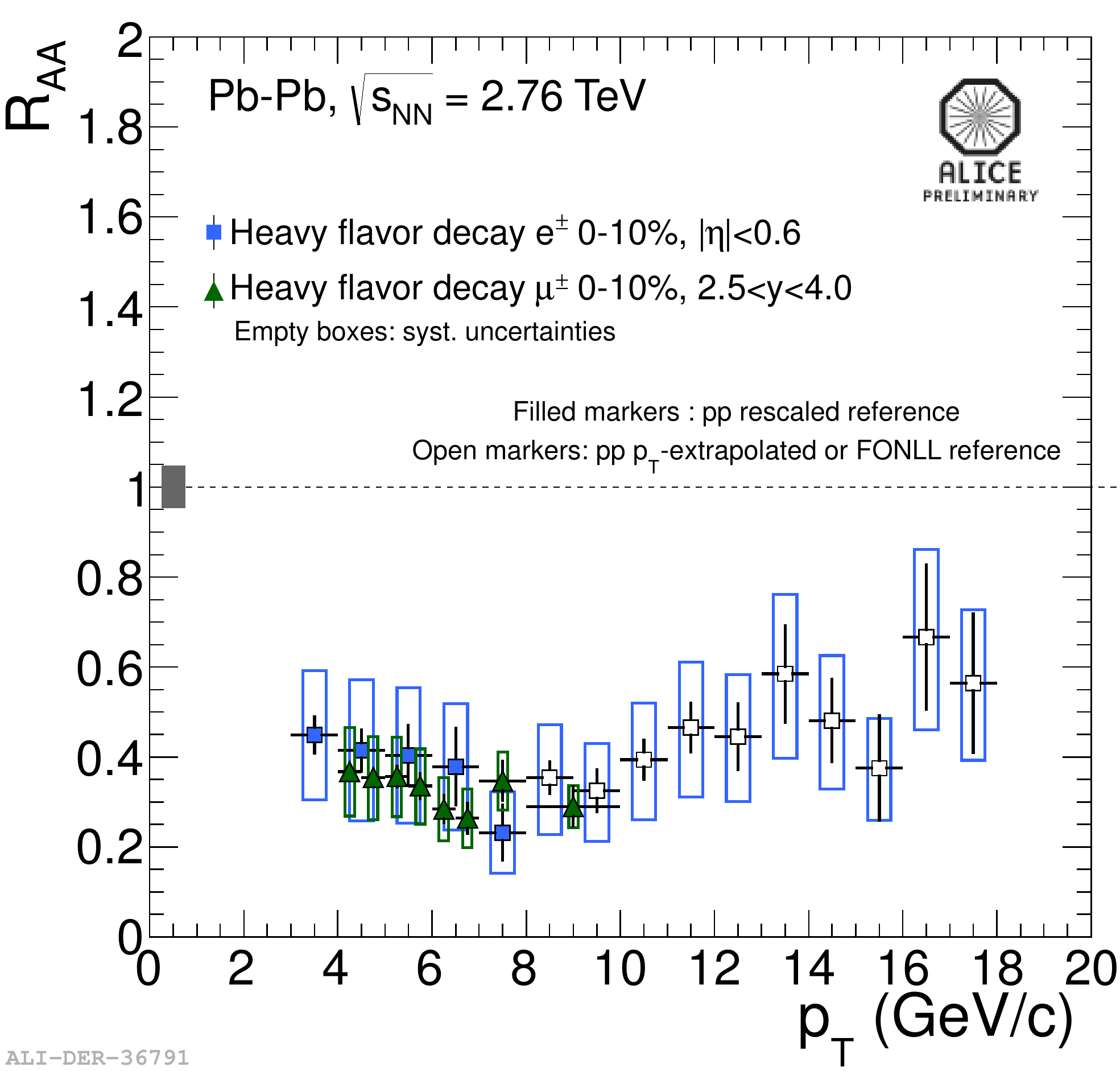}
  \caption{Nuclear modification factors for heavy-flavour decay muons at forward rapidity~\cite{hfmraa} and electrons 
  at central rapidity in central (0--10\%) Pb--Pb collisions, as a function of $\pt$.}
\label{fig:leptRAA}
\end{center}
\end{figure}

\begin{figure}[!t]
  \begin{center}
\includegraphics[width=0.8\textwidth]{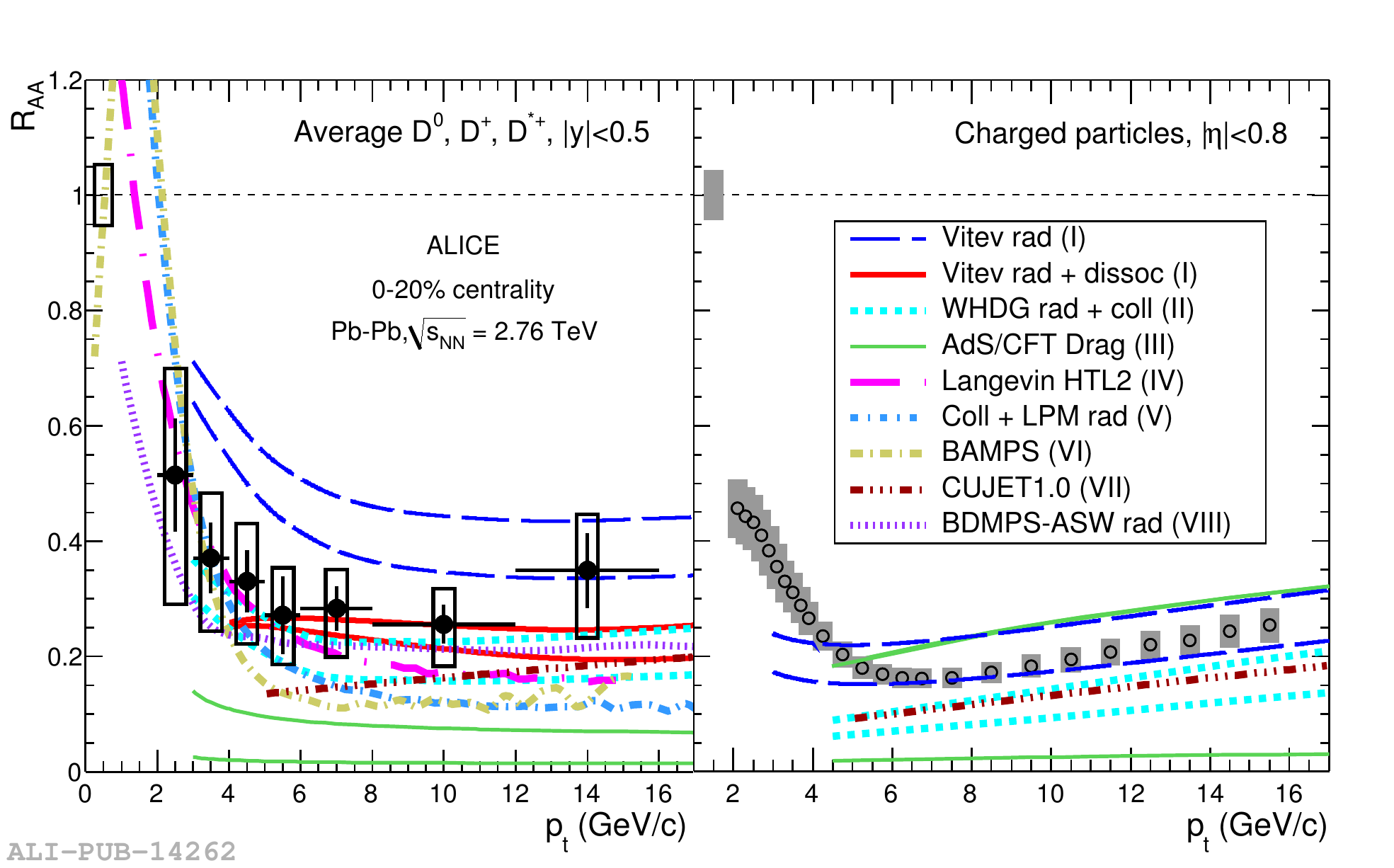}
  \caption{Average $\RAA$ of D mesons (left) and $\RAA$ of charged particles (right) in the \mbox{0--20\%} centrality class compared to model calculations (details in the text)~\cite{aliceDRAA}.}
\label{fig:models}
\end{center}
\end{figure}

A large suppression is also observed for heavy-flavour decay muons~\cite{hfmraa} and electrons, 
at forward and central rapidity, respectively, as shown for the centrality class 0--10\% in figure~\ref{fig:leptRAA}. 
Both lepton species exhibit a suppression of factor of about 3 up to $10~\gev/c$, and about 2 for electrons up to $\pt=18~\gev/c$
in a momentum range where beauty decays should be dominant.

Several theoretical models based on parton energy loss compute the heavy-flavour 
nuclear modification factor: (I)~\cite{vitev,vitevjet}, (II)~\cite{whdg2011}, (III)~\cite{horowitzAdSCFT}, (IV)~\cite{beraudo}, (V)~\cite{gossiaux}, (VI)~\cite{bamps}, (VII)~\cite{cujet}, (VIII)~\cite{adsw}. Figure~\ref{fig:models} displays the comparison of these
models  
to the average D meson $\RAA$, for central Pb--Pb collisions (0--20\%), along with the 
comparison to the charged-particle $\RAA$~\cite{RAAcharged}, for those models that also compute 
this observable. Radiative energy loss supplemented with in-medium D meson dissociation (I)~\cite{vitev}
and radiative plus collisional energy loss in the WHDG (II)~\cite{whdg2011} and CUJET1.0 (VII)~\cite{cujet} implementations describe reasonably well at the same time the charm and 
light-flavour suppression. 
A model based on AdS/CFT drag coefficients (III)~\cite{horowitzAdSCFT} underestimates significantly the charm $\RAA$ and
has very limited predictive power for the light-flavour $\RAA$.

\vspace{-0.2cm}

\section{Conclusions}

The ALICE results on the $\RAA$ 
nuclear modification factors of heavy-flavour
hadrons in Pb--Pb collisions at the LHC imply a strong 
in-medium energy loss for c and b quarks.
The $\rm D^0$, $\rm D^+$ and $\rm D^{*+}$ $\RAA$, 
measured as a function of $\pt$ and centrality,
is as low as 0.2--0.3 for $\pt>5~\gev/c$ and compatible with the $\RAA$ of pions.
Below $5~\gev/c$, there is a hint for $\RAA^{\rm D}>\RAA^\pi$. 
The comparison with J/$\psi$ from B decays (CMS Collaboration) reveals a first indication 
for different suppression of charm and beauty hadrons in central collisions.
A large suppression, $\RAA\approx 0.3$--0.5
is observed also for heavy-flavour decay electrons and muons in 
the $\pt$ range above $5~\gev/c$, where beauty decays are dominant according to 
pQCD calculations. 
The completion of the analysis of high-statistics data from the 2011 Pb--Pb run should allow 
for sharper conclusions on the comparison of the D meson $\RAA$ with pions and with 
beauty measurements. For the latter, the separation of electrons from B decays should also be possible.
Finally, data from the p--Pb run scheduled for January 2013 will constrain the relevance
of initial-state nuclear effects in the low $\pt$ range.


\begin{thebibliography}{10}

\bibitem{aliceJINST} 
 K.~Aamodt {\it et al.} [ALICE Collaboration], JINST \textbf{3} (2008) S08002.

\bibitem{glauber} 
M. Miller, K. Reygers, S. Sanders, and P. Steinberg, Ann. Rev. Nucl. Part. Sci. {\bf 57} (2007) 205.

\bibitem{dk}
 Y.~L.~Dokshitzer, D.~E.~Kharzeev,
  Phys.\ Lett.\  {\bf B519 } (2001)  199-206.
  
\bibitem{asw}
  N.~Armesto, C.~A.~Salgado, U.~A.~Wiedemann,
  Phys.\ Rev.\  {\bf D69} (2004)  114003.

\bibitem{whdg}
  S.~Wicks, W.~Horowitz, M.~Djordjevic, M.~Gyulassy,
  Nucl.\ Phys.\  {\bf A783 } (2007)  493-496.

\bibitem{rapp}
  H.~van Hees, V.~Greco and R.~Rapp,
  Phys.\ Rev.\ {\bf C73} (2006) 034913.

\bibitem{adsw}
  N.~Armesto, A.~Dainese, C.~A.~Salgado, U.~A.~Wiedemann,
  Phys.\ Rev.\  {\bf D71 } (2005)  054027.


\bibitem{aliceDRAA}
  B.~Abelev {\it et al.}  [ALICE Collaboration],
  JHEP {\bf 1209} (2012) 112.

\bibitem{fonllpriv}
  M.~Cacciari {\it et al.},
  JHEP {\bf 1210} (2012) 137.
  
\bibitem{hfepp7}
  B.~Abelev {\it et al.}  [ALICE Collaboration],
  arXiv:1205.5423 [hep-ex].  
  
  
  \bibitem{hfmpp7}
  B.~Abelev {\it et al.}  [ALICE Collaboration],
  Phys.\ Lett.\ {\bf B708} (2012) 265.


\bibitem{hfmraa}
  B.~Abelev {\it et al.}  [ALICE Collaboration],
  PRL {\bf 109} (2012) 112301.





\bibitem{Dpp2.76}
  B.~Abelev {\it et al.}  [ALICE Collaboration],
  JHEP {\bf 1207} (2012) 191.

\bibitem{Dpp7}
  B.~Abelev {\it et al.}  [ALICE Collaboration],
  JHEP {\bf 1201} (2012) 128.






\bibitem{scaling}
R. Averbeck, N. Bastid, Z. Conesa del Valle, A. Dainese, X. Zhang, arXiv:11073243 (2011).

\bibitem{RAAcharged}
  B.~Abelev {\it et al.}  [ALICE Collaboration],
  arXiv:1208.2711 [hep-ex].

\bibitem{CMSquarkonia}
CMS Collaboration, arXiv:1201.5069 [nucl-ex] (2012).

\bibitem{CMSpashin12014}
CMS Collaboration,  CMS-PAS-HIN-12-014 (2012).


\bibitem{vitev}
  R.~Sharma, I.~Vitev and B.~-W.~Zhang,
  Phys.\ Rev.\  {\bf C80 } (2009)  054902.

\bibitem{vitevjet}
  Y.~He, I.~Vitev and B.~-W.~Zhang,
  Phys.\ Lett.\ {\bf B713} (2012) 224.


\bibitem{whdg2011}
  W.~A.~Horowitz and M.~Gyulassy,
  J.\ Phys.\ {\bf G38} (2011) 124114.



\bibitem{horowitzAdSCFT} 
  W.~A.~Horowitz,
  arXiv:1108.5876 [hep-ph] (2011).
  
\bibitem{beraudo}
  W.~M.~Alberico, {\it et al.},
  Eur.\ Phys.\ J.\ {\bf C71} (2011) 1666;
  J.\ Phys.\ {\bf G38} (2011) 124144.
  
\bibitem{gossiaux}
  P.~B.~Gossiaux, J.~Aichelin, T.~Gousset and V.~Guiho,
  J.\ Phys.\ {\bf G37} (2010) 094019.

  \bibitem{bamps}
  O.~Fochler, J.~Uphoff, Z.~Xu and C.~Greiner,
  J.\ Phys.\ {\bf G38} (2011) 124152.
  
  
\bibitem{cujet} 
A.~Buzzatti and M.~Gyulassy,
  Phys.\ Rev.\ Lett.\  {\bf 108} (2012) 022301.




\end{thebibliography}
\end{document}